\documentclass{Interspeech}
\usepackage{hyperref}
\title{ViP-VL: Vietnamese Self-supervised Speech Pretraining Model with Vector-Quantization Learning}
\author[affiliation={1}, orcid=0009-0003-1190-5569, equalcontribution]{Khanh}{Le}
\author[affiliation={2}, orcid=0000-0001-8438-6585, equalcontribution]{Kiet Anh}{Hoang}
\author[affiliation={2}, orcid=0009-0009-4515-7056, equalcontribution]{Bao}{Nguyen}
\author[affiliation={2}, orcid=0009-0006-9774-594X, equalcontribution]{Duy}{Vo}
\author[orcid=0009-0001-1973-2346, affiliation={2}]{Dung}{Vo}
\author[orcid=0009-0002-1422-9685, affiliation={2}]{\\ Thai}{Tran}
\author[orcid=0009-0003-1560-2813, affiliation={2}]{Linh}{Pham}
\author[orcid=0000-0002-1610-8206, affiliation={1}]{Khoa}{D Doan}

\address{
    $^1$ VinUniversity, Vietnam \\
    $^2$ UNEY, Switzerland 
}

\email{\{khanhld218, doankhoadang\}@gmail.com, \{kiet.ha, bao.ld, duy.vhb\}@uney.com}

\keywords{ssl, vietnamese, chunkformer, vip-vl}

\usepackage{comment}
\usepackage{booktabs}
\usepackage{soul}

\begin{document}
\maketitle

\begin{abstract}

We present ViP-VL, an efficient \textbf{Vi}etnamese Self-supervised speech \textbf{P}retraining model leveraging \textbf{V}ector-quantization \textbf{L}earning. To bridge the gap between high-resolution audio and efficient processing, ViP-VL incorporates Acoustic Stacking and Receptive Field Alignment to enable a synchronized 8x subsampling rate within the ChunkFormer architecture, while further enhancing representation robustness through a specialized Mask Selection Strategy during pretraining on the BEST-RQ framework. Pretrained on 17,000 hours of unlabeled Vietnamese speech, our model establishes new state-of-the-art results across four major downstream tasks: Automatic Speech Recognition, Speech Emotion Recognition, Dialect Classification, and Speaker Verification. To facilitate future research and the development of high-performance Vietnamese speech technologies, we publicly release our pretrained weights and implementation at \href{https://github.com/khanld/chunkformer}{\texttt{github.com/khanld/chunkformer}}.

\end{abstract}
\sloppy
\sloppy

\section{Introduction}

Self-supervised learning (SSL) has recently driven significant advancements in speech processing. By leveraging vast amounts of unlabeled data, these approaches enable the models to learn robust acoustic representations that, when combined with supervised fine-tuning, substantially improve performance. This capability is particularly advantageous for low-resource languages such as Vietnamese, where annotated data is scarce, offering a viable pathway to bridge the performance gap between high-resource and underserved linguistic communities. 


\vspace{2pt} \noindent \textbf{Speech SSL.} Modern speech SSL generally falls into two categories: contrastive and predictive. Contrastive approaches, exemplified by wav2vec 2.0 \cite{baevski2020wav2vec}, quantize speech features into a latent codebook and utilize a contrastive loss to distinguish positive targets from distractors. 
Wav2vec-C \cite{sadhu21_interspeech}, a later extension, incorporates a consistency loss for reconstruction, while XLS-R \cite{babu22_interspeech} demonstrates the benefits of massive multilingual pretraining. 
In contrast, predictive approaches, pioneered by HuBERT \cite{hsu2021hubert}, treat pretraining as a masked token prediction task by generating discrete targets via $k$-means clustering on intermediate features.
W2v-BERT \cite{chung2021w2v} subsequently unified these paradigms by utilizing contrastive and predictive losses simultaneously. 
More recently, BEST-RQ \cite{chiu2022self} simplified this pipeline by replacing the expensive clustering with a fixed random-projection quantization.
This method streamlines training while achieving comparable, and sometimes even better, performance than that of the complex clustering-based models, making it an appealing option for low-resource languages. Nevertheless, a significant hurdle in its adoption is that public implementations~\cite{Whetten2024OpenIA, speechbrain_v1, zhang2022wenet} fail to reproduce the high performance reported in the original benchmarks, especially in low-resource contexts.

Another major barrier of existing prevalent SSL approaches is their computational burden in both inference speed and memory, largely due to backbones with 20ms frame shift~\cite{baevski2020wav2vec, hsu2021hubert, chen2022wavlm}. While BEST-RQ improves efficiency with a 40ms frame length, further optimizations are necessary to accelerate the pretraining phase and to meet the practical demands of real-time applications and large-scale systems. Recent frameworks like NEST~\cite{huang2025nest} have integrated a FastConformer~\cite{rekesh2023fast} backbone with $8\times$ subsampling, resulting in an 80ms frame length that significantly increases throughput. However, these works often bypass thorough methodological verification, relying instead on the ``brute force'' of large-scale pretraining. Consequently, it remains unclear whether their gains stem from architectural design or simply data scale. While subsampling enhances efficiency, excessive rates lead to temporal sparsity and a loss of fine-grained acoustic resolution. This results in representation instability and increased error rates, as the model can no longer capture the short-duration phonetic cues essential for robust speech processing \cite{mfa-conformer, 10.5555/3600270.3600950, xu-etal-2024-conformer,e24070866}.

\vspace{2pt} \noindent \textbf{Vietnamese Speech SSL.} While the first Vietnamese SSL model \cite{Thai_Binh_Nguyen_wav2vec2_vi_2021} successfully utilized 13k hours of unlabeled audio, it inherited the high inference latency inherent to the wav2vec 2.0 architecture. Subsequent large-scale efforts like VietASR~\cite{zhuo25_interspeech} demonstrated the potential of 70,000-hour pretraining but withheld public weights. This leaves a critical gap: the Vietnamese ecosystem lacks a high-performance SSL model that is both computationally optimized for deployment and publicly accessible to the community. 

\vspace{2pt} \noindent \textbf{Contributions.} This work addresses these challenges by systematically investigating the mechanics of SSL within an $8\times$ subsampling architecture. We find that a correct synchronization between the masking manifold and the encoder's subsampling rate is crucial for performance. Based on this insight, we extend the BEST-RQ framework  with a ChunkFormer encoder and correct its synchronization problem with acoustic stacking, receptive field alignment, and an optimized masking strategy. The proposed implementation, called ViP-VL (\textbf{Vi}etnamese self-supervised speech \textbf{P}retraining via \textbf{V}ector-quantization \textbf{L}earning), is pretrained on 17,000 hours of Vietnamese audio and has 78M-parameters.  Extensive evaluations demonstrate the state-of-the-art performance and computational efficiency of ViP-VL across four Vietnamese speech benchmarks: Automatic Speech Recognition, Speech Emotion Recognition, Dialect Classification, and Speaker Verification. 

\section{ViP-VL}
\subsection{Architecture}\label{sec:architecture}
ViP-VL leverages BEST-RQ, a paradigm that streamlines self-supervised learning via a frozen, randomly initialized quantizer. This approach eliminates the need for the computationally expensive codebook training required by wav2vec 2.0~\cite{baevski2020wav2vec} or the iterative clustering used in HuBERT~\cite{hsu2021hubert}. By utilizing fixed random projections to map input speech signals to discrete indices, BEST-RQ enables Masked Language Modeling without a learned tokenizer. This significantly simplifies the training pipeline and, as demonstrated in prior work, achieves a $2.4\times$ speedup in pre-training compared to wav2vec 2.0. Furthermore, we adopt ChunkFormer~\cite{le2025chunkformer} as the encoder's backbone for its high efficiency and robustness.
The key architectural feature of ChunkFormer is its chunk-wise processing strategy. Instead of full-sequence attention, ChunkFormer employs relative right-context attention, allowing each chunk to attend to a limited number of future frames across chunk boundaries via relative positional encodings. This mitigates the global boundary effects while keeping the attention computation local and efficient. ViP-VL employs an aggressive $8\times$ temporal subsampling stage, substantially reducing the sequence length processed by the self-attention layers. To compensate for the loss of temporal resolution, the convolutional modules adopt the kernel sizes of 15 and increased channel capacity, yielding a receptive field comparable to that of a full-resolution Conformer encoder~\cite{gulati20_interspeech}.

\subsection{Training Objective}
We construct the quantizer input $\mathbf{x}_t \in \mathbb{R}^{d}$ by stacking adjacent acoustic frames (e.g., log-Mel filterbanks) with a stride corresponding to the encoder's subsampling rate. We first apply input mean-variance normalization to $\mathbf{x}_t$ to align the feature distribution with the standard normal codebook. This step is critical to prevent codebook collapse and to promote uniform codebook utilization~\cite{chiu2022self}. We then project $\mathbf{x}_t$ using a frozen random matrix $\mathbf{A} \in \mathbb{R}^{h \times d}$ initialized via Xavier initialization, and select the nearest neighbor in a codebook $\mathcal{C} = \{\mathbf{c}_1, \ldots, \mathbf{c}_n\}$ sampled from a standard normal distribution $\mathbf{c}_i \sim \mathcal{N}(0, \mathbf{I})$ with $\mathbf{c}_i \in \mathbb{R}^{h}$. Formally, the quantizer assigns a discrete label $y_t$ by
\begin{equation}
  y_t = \operatorname*{argmin}_{i} \left\lVert \frac{\mathbf{c}_i}{\lVert\mathbf{c}_i\rVert_2} - \frac{\mathbf{A}\mathbf{x}_t}{\lVert\mathbf{A}\mathbf{x}_t\rVert_2} \right\rVert_2.
\end{equation}
Leveraging these discrete targets, the pretraining objective follows the masked language modeling paradigm of BERT~\cite{devlin2019bert}, adapted to continuous speech~\cite{chiu2022self}. Let $\mathbf{X} = (\mathbf{x}_1, \ldots, \mathbf{x}_T)$ denote the original feature sequence, and let $\mathcal{M} \subset \{1,\ldots,T\}$ be the set of masked time indices sampled using span masking.
We construct a corrupted sequence $\tilde{\mathbf{X}} = (\tilde{\mathbf{x}}_1, \dots, \tilde{\mathbf{x}}_T)$ by replacing $\mathbf{x}_t$ with a learnable mask embedding $\mathbf{E}_{\text{mask}} \in \mathbb{R}^d$, initialized from a uniform distribution $\mathcal{U}(0, 1)$ for all $t \in \mathcal{M}$, while keeping $\tilde{\mathbf{x}}_t = \mathbf{x}_t$ for $t \notin \mathcal{M}$.
The encoder maps $\tilde{\mathbf{X}}$ to contextual representations, which are then projected to predict the discrete targets $\{y_t\}_{t \in \mathcal{M}}$ computed from the original unmasked sequence $\mathbf{X}$. The model is trained by minimizing the average negative log-likelihood over masked positions:
\begin{equation}
  \mathcal{L}(\theta) = - \frac{1}{|\mathcal{M}|} \sum_{t \in \mathcal{M}} \log P_{\theta}(y_t \mid \tilde{\mathbf{X}}),
\end{equation}
where $P_{\theta}(y_t \mid \tilde{\mathbf{X}})$ denotes the predicted probability assigned to the true codebook index $y_t$.

\subsection{Acoustic Stacking and Receptive Field Alignment} 
A critical component of our implementation is the synchronization between the masking manifold and the encoder's subsampling rate. In contrast to Nest \cite{huang2025nest}, which concatenates frames to match the stride while overlooking internal padding and kernel dynamics, our strategy uses a stacking window of 15 frames with a stride of 8. This configuration is mathematically derived to mirror the receptive field of the encoder’s input stage, which consists of 3-stacked convolutions with a kernel size of 3 and stride of 2. This precise alignment ensures that unmasked features are temporally synchronized with the encoder’s output manifold. For feature aggregation, we evaluate both concatenated stacking and average stacking. While averaging reduces input dimensionality for the random projection layer, it functions as a low-pass filter that smooths fine-grained acoustic variations. Empirical observations indicate that concatenated stacking yields superior performance by preserving the full information density and feature variance of the local receptive field. This provides the random projection layer with more distinctive acoustic cues, resulting in more diverse codebook utilization.

\subsection{Mask Selection Strategy}
A fundamental design choice in self-supervised learning is whether to apply the masking manifold before or after the temporal subsampling module. While studies on Wav2vec2 \cite{baevski2020wav2vec} or Wav2vec2-Conformer \cite{zhang2022pushinglimitssemisupervisedlearning} architectures perform masking post-subsampling to reduce the implementation complexity, our empirical evaluations favor masking the raw 10ms filterbank frames prior to subsampling. However, this introduces a mapping complexity: a single subsampled frame corresponds to a receptive field of 15 adjacent 10ms frames. To bridge this temporal resolution gap, we propose a probabilistic masking threshold: a subsampled frame is designated as ``masked'' if and only if at least 80\% (i.e., 12 out of 15) of the constituent pre-subsampled frames are masked. This strict threshold prevents the model from exploiting local acoustic leakages from partially unmasked frames, maintaining a high level of task difficulty. With a mask length of 400ms and a 0.01 masking probability, this strategy yields an effective time masking ratio of 45\% during training.


\section{Experiments}
\subsection{ViP-VL Pretraining and Evaluations}
\vspace{2pt}\textbf{Proposal Verification.} We first validate our architecture by pretraining on the 960-hour LibriSpeech dataset \cite{panayotov2015librispeech} and fine-tuning on the 100-hour subset. As shown in Table~\ref{tab:librispeech_sanity}, our method bridges the performance gap typical of high compression, achieving performance comparable to the $2\times$ baseline while reducing the self-attention computation by 16 times. By outperforming $8 \times$ subsampling BEST-RQ without alignment or masking, we demonstrate that architectural precision is as critical as data scale for high-compression SSL. 
\begin{table}[h]
\centering
  \caption{LibriSpeech verification results. Lower WER is better. \textbf{Bold} and \underline{underline} indicate the best and second-best results, respectively. This convention applies to all subsequent tables.}
  \label{tab:librispeech_sanity}
  \setlength{\tabcolsep}{6pt}
  \renewcommand{\arraystretch}{1.15}
  \begin{tabular}{lccc}
    \toprule
    \textbf{Model} & \textbf{test-clean} & \textbf{test-other} & \textbf{avg.} $\downarrow$ \\
    \midrule
    $8 \times$ BEST-RQ \cite{chiu2022self} & 6.8 & 17.0 & 11.9 \\
    wav2vec 2.0~\cite{baevski2020wav2vec} & \underline{6.1} & \textbf{13.3} & \textbf{9.7} \\
    \midrule

    \textbf{ViP-VL (ours)}  & \textbf{5.3} & \underline{14.1} & \textbf{9.7} \\
    \bottomrule
  \end{tabular}
    \vspace{-6pt}

\end{table}

\vspace{2pt} \noindent \textbf{Pretraining.} The pretraining of ViP-VL uses a diverse Vietnamese corpus totaling approximately 17,000 hours of unlabeled speech. The corpus aggregates multiple multi-domain sources to ensure acoustic and linguistic variety, including GigaSpeech 2 \cite{yang-etal-2025-gigaspeech}, the MSR-86K \cite{msr} corpus, and other public-domain sources.
Our ViP-VL model comprises 12 blocks with a 512-dimension output size, 8 attention heads, and 2,048 linear units in the position-wise feed-forward layers, totaling 78M parameters. Our BEST-RQ setup uses a codebook size of 1,024 and a 16-dimensional projection space. Pretraining is conducted on 8$\times$ NVIDIA H200 GPUs for 320,000 steps. The final pretrained checkpoint is obtained by averaging the weights of the last 50 epochs. 

\vspace{2pt} \noindent \textbf{Evaluation Setup.} We benchmark our pretrained model across four downstream tasks: Automatic Speech Recognition (ASR), Speech Emotion Recognition (SER), Speech Dialect Classification (SDC), and Speaker Verification (SV). We focus exclusively on models following the self-supervised or weakly-supervised pretraining and supervised fine-tuning paradigm, rather than models trained on massive proprietary supervised datasets. Except for the speaker verification task, well-known ECAPA-TDNN~\cite{ecapa} and ResNet34~\cite{zeinali2019but} architectures are also used for evaluation.
The selected baselines encompass the following architectures:
\begin{itemize}
    \item \textbf{PhoWhisper (Base \& Large \cite{PhoWhisper}):} Utilizing the multilingual Whisper backbone \cite{radford2022robustspeechrecognitionlargescale}, these are the current state-of-the-art for Vietnamese, pretrained on 680k hours of weak-supervised data.
    \item \textbf{Wav2vec2-Vi (Base \& Large \cite{Thai_Binh_Nguyen_wav2vec2_vi_2021}):} Based on the Wav2vec 2.0 architecture \cite{baevski2020wav2vec}, pretrained on 13k hours of unlabeled Vietnamese YouTube audio.
    \item \textbf{VietASR \cite{zhuo25_interspeech}:} An RNN-T model, pretrained on 70,000 hours of audio following the HuBERT self-supervised learning paradigm.
\end{itemize}
To evaluate the robustness of our pretraining and ensure there is no data leakage, all fine-tuning datasets listed in Table \ref{tab:datasets} are strictly excluded from our pretraining corpus. This protocol ensures that the model's performance on the benchmarks is a result of its ability to generalize to unseen acoustic distributions and domains, rather than mere exposure to the target data during the self-supervised phase.
\begin{table}[h]
\centering
\caption{Summary of Fine-tuning Datasets}
\label{tab:datasets}
\begin{tabular}{@{}lcc@{}}
\toprule
\textbf{Task} & \textbf{Dataset} & \textbf{Duration} \\
\midrule
ASR & VLSP 2020\footnotemark & 250h \\
SER & ViSEC \cite{viSEC_icassp_2024} & 3h \\
SDC & ViMD \cite{dinh-etal-2024-multi} & 102h \\
SV & VoxVietnam \cite{vu2025voxvietnam} & 261h (1,406 speakers) \\
\bottomrule
\end{tabular}
\vspace{-5pt}
\end{table}
\footnotetext{\url{https://vlsp.org.vn/vlsp2020/eval/asr}}
\subsection{Automatic Speech Recognition}
\label{sec:results:asr}
Fine-tuning is conducted over 150 epochs across 4$\times$ NVIDIA H200 GPUs, optimized via the CTC loss function. We employ the AdamW optimizer with a peak learning rate of $5 \times 10^{-5}$. To stabilize the early stages of training, a linear warmup of 10,000 steps is applied.

As shown in Table~\ref{tab:asr_benchmarks}, the proposed ViP-VL achieves state-of-the-art average performance across all evaluated Vietnamese ASR benchmarks. ViP-VL has the lowest average WER (13.76\%), outperforming larger models such as Wav2vec2-Large-Vi and the data-intensive PhoWhisper-Large. Given that all models were fine-tuned using the VLSP 2020 dataset (or a subset thereof in the case of PhoWhisper), this comparison provides a direct assessment of each architecture’s capability and the effectiveness of our pre-training strategy. Although our results are superior to those of VietASR, this comparison remains primarily qualitative due to disparities in fine-tuning data size. The benchmark results are obtained using the publicly released semi-supervised VietASR model finetuned on 70,000 hours of Vietnamese speech data. These results highlight that ViP-VL not only excels in parameter efficiency but also offers superior generalization capabilities across diverse datasets.

\begin{figure}[t]
    \centering
    \includegraphics[width=\linewidth]{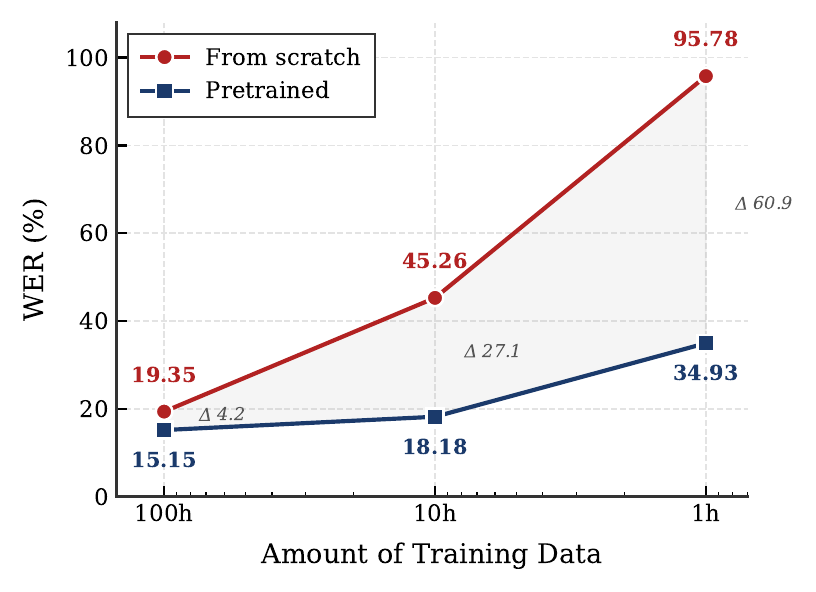}
    \vspace{-15pt}
    \caption{Word error rate (WER) comparison between from scratch and pretrained on VLSP-T1.}
    \label{fig:wer_comparison}
    \vspace{-9pt}
\end{figure}

\vspace{2pt} \noindent \textbf{Effectiveness of SSL Pretraining.} The effectiveness of SSL pretraining on low-resource ASR tasks is illustrated in Figure \ref{fig:wer_comparison}, which highlights a significant performance gap between models trained from scratch and those using self-supervised pretraining. As the amount of labeled training data decreases, the benefits of pretraining become increasingly pronounced.
\begin{table*}[t]

    \centering
    \caption{Comparison with prior Vietnamese speech pretraining models on ASR benchmarks.}
    \label{tab:asr_benchmarks}
    \setlength{\tabcolsep}{5.3pt} 
    \begin{tabular}{lcccccccccc}
        \toprule
        \textbf{Model} & \textbf{Params} & \textbf{Pretrain} & \textbf{Finetune} & \textbf{Head} & \textbf{VIMD} & \textbf{VLSP-T1} & \textbf{VLSP-T2} & \textbf{VIVOS} & \textbf{Giga.} & \textbf{Avg} $\downarrow$ \\
        \midrule
        Wav2vec2-Base-Vi  & 95M   & 13,000h & 250h & CTC  & 15.63 & 16.82 & 44.91 &  9.90 & 16.74 & 20.80 \\   
        Wav2vec2-Large-Vi & 317M  & 13,000h & 250h & CTC  & 14.45 & 15.18 & 36.75 &  8.61 & 14.47 & 17.89 \\
        PhoWhisper-Base   & 74M   & 680,000h & 800h & AED  & 19.77 & 19.70 & 43.01 &  8.46 & 20.75 & 22.34 \\
        PhoWhisper-Large  & 1.55B & 680,000h & 800h & AED  & 12.74 & \underline{13.75} & \textbf{26.68} &  \textbf{4.67} & 12.60 & \underline{14.09} \\
        VietASR           & 68M   & 70,000h & 70,000h & RNN-T & \textbf{9.92} & 14.47 & 34.78 & 7.21 & \textbf{7.69} & 14.81 \\
        \midrule
        \textbf{ViP-VL (ours)}   & 78M   & 17,000h & 250h & CTC   & \underline{10.91} & \textbf{11.20} & \underline{31.61} & \underline{5.25} & \underline{9.85} & \textbf{13.76} \\
        \bottomrule
    \end{tabular}
        \vspace{-5pt}

\end{table*}
\subsection{Speech Emotion Recognition}
The ViSEC dataset contains four emotion categories: neutral, happy, sad, and angry. Following the ViSEC partitioning protocol, we utilize 5-fold cross-validation to ensure robust evaluation across all segments. For supervised fine-tuning, the last-layer features are passed through a simple classifier head consisting of an average pooling layer followed by a 256-unit hidden layer with a dropout rate of 0.1. All models are trained with a batch size of 16, with training terminated with early stopping (patience of 10 epochs). We utilize the AdamW optimizer with an initial learning rate of $10^{-5}$ and a weight decay of $10^{-6}$. The schedule is further refined with a 500-step linear warmup and a subsequent linear decay reaching a floor of 1\% of the initial rate. The Unweighted Accuracy (UA) is used for evaluation.

Table \ref{tab:merged_results_clean} shows the performance of different pretrained models on the ViSEC dataset. Our model significantly outperforms all baselines, achieving an UA of 74.45\%. Compared to the strongest baseline, Wav2vec2-Large-Vi, ViP-VL achieves an absolute improvement of 1.45\% score. Furthermore, ViP-VL exhibits low performance variance across runs, as indicated by smaller standard deviations, highlighting its stability and reliability. These results confirm that the proposed approach establishes new state-of-the-art on the ViSEC dataset. 


\subsection{Speech Dialect Classification}
We categorize the dialect classification task into two levels of granularity: regional and provincial. The regional sub-task labels utterances into three broad classes (Northern, Central, and Southern), whereas the provincial sub-task demands a more fine-grained classification across 63 distinct provinces. Consistent with our Speech Emotion Recognition (SER) framework, we maintain the same training configuration and employ the F1-score as our primary evaluation metric. Results are summarized in Table~\ref{tab:merged_results_clean}. For regional classification, ViP-VL achieves the highest performance with an F1-score of 93.24\%. While provincial classification proves significantly more challenging due to its higher granularity, ViP-VL still maintains its lead, reaching a top score of 57.17\% and outperforming the second-best model, PhoWhisper-Large (54.91\%).

\begin{table}[ht!]
  \centering
  \caption{Performance comparison across different tasks. Values represent average $\pm$ standard deviation where available.}
  \label{tab:merged_results_clean}
    \setlength{\tabcolsep}{4pt} 
  \begin{tabular}{lccc}
    \toprule
    \textbf{Model} & \textbf{Emotion} $\uparrow$ & \textbf{Region} $\uparrow$ & \textbf{Province} $\uparrow$ \\
    \midrule
    Wav2vec2-Base-Vi  & $71.79 \pm 1.01$ & 91.57 & 41.12 \\
    Wav2vec2-Large-Vi & \underline{$73.00 \pm 1.72$} & \underline{92.15} & \underline{54.91} \\
    PhoWhisper-Base   & $70.92 \pm 2.44$ & 87.14 & 39.53 \\
    PhoWhisper-Large  & $72.68 \pm 2.90$ & 90.14 & 49.67 \\
    \midrule
    \textbf{ViP-VL (ours)} & \textbf{74.45 $\pm$ 1.05} & \textbf{93.24} & \textbf{57.17} \\
    \bottomrule
  \end{tabular}
  \vspace{-5pt}
\end{table}

\subsection{Speaker Verification}
The \textbf{VoxVietnam-T} and \textbf{VoxVietnam-O} sets are utilized for training and evaluation, respectively. During the training phase, audio samples are randomly cropped into 2-second segments. We employ data augmentation using the MUSAN noise corpus~\cite{snyder2015musan} and RIR reverberation~\cite{ko2017study} with a probability of 0.6. The model is optimized using AAM-Softmax~\cite{deng2019arcface} with a scale $s=30$ and margin $m=0.2$. We utilize a cosine learning rate scheduler with a peak learning rate of $10^{-3}$ and a 5-epoch warmup. SSL-based models are trained for a total of 30 epochs and the SSL encoder remains frozen for the first 20 epochs before proceeding with full-model fine-tuning. Other models are trained from scratch for 100 epochs.
For SSL models (ViP-VL, Wav2vec2-Base-Vi and Wav2vec2-Large-Vi), we extract the hidden states from all layers and utilize a weighted average followed by linear layers for feature extraction. All models utilize Attentive Statistics Pooling~\cite{okabe18_interspeech} for temporal aggregation.

For evaluation, speaker embeddings are scored using cosine similarity. Performance is reported via Equal Error Rate (EER) and the minimum normalized Detection Cost Function (minDCF) with $P_{\text{target}} = 0.05$ and $C_{\text{fa}} = C_{\text{miss}} = 1$. All experiments are conducted using the WeSpeaker toolkit~\cite{wang2023wespeaker}.

Table~\ref{tab:sv_result} presents a comparative analysis of the proposed ViP-VL against several competitive baselines on the VoxVietnam-O set. Our model achieves the lowest EER of~\textbf{3.639\%}, outperforming both supervised backbones and the \mbox{Wav2vec2-Base-Vi} SSL model. Although Wav2vec2-Large-Vi yields a slightly superior minDCF of 0.504, ViP-VL remains highly competitive with a minDCF of 0.518 while maintaining a significantly lower EER than the Large variant. Interestingly, the \mbox{Wav2vec2-Large-Vi} model exhibits performance degradation in terms of EER compared to the Base variant. This suggests that the increased parameter capacity of the large architecture may lead to overfitting on the speaker-specific nuances of the VoxVietnam corpus, whereas ViP-VL maintains a more robust and generalizable representation. These results indicate that ViP-VL provides a more balanced and discriminative embedding space for Vietnamese speaker verification.
\begin{table}[th]
  \caption{Speaker verification performance on VoxVietnam-O. EER (\%) and minDCF are reported (lower is better).}
  \label{tab:sv_result}
  \centering
  \begin{tabular}{lcc}
    \toprule
    \textbf{Model} & \textbf{EER (\%)} $\downarrow$ & \textbf{minDCF} $\downarrow$ \\
    \midrule
    ECAPA-TDNN        & 3.925 & 0.573 \\
    ResNet34          & 4.007 & 0.567 \\
    Wav2vec2-Base-Vi  & \underline{3.679} & 0.523 \\
    Wav2vec2-Large-Vi & 4.334 & \textbf{0.504} \\
    \midrule
    \textbf{ViP-VL (ours)} & \textbf{3.639} & \underline{0.518} \\
    \bottomrule
  \end{tabular}
  \vspace{-5pt}
\end{table}

\section{Conclusion}

In this paper, we introduce ViP-VL, an efficient Vietnamese self-supervised speech pretraining model leveraging Vector-quantization Learning. By combining the BEST-RQ framework with a ChunkFormer encoder, a receptive field-aligned stacking strategy, and a specialized mask selection strategy, we achieve state-of-the-art performance across multiple Vietnamese benchmarks. Our results demonstrate that ViP-VL provides a computationally accessible yet high-performance solution specifically tailored for the Vietnamese language. This work motivates future development of low-resource SSL speech models, including compressed small-scale model variants of ViP-VL, knowledge distillation for enhanced model efficiency, and expansion of the pretraining corpus, as well as extensions in other low-resource languages.


\section{Generative AI Use Disclosure}
Generative AI tools were used for editing and polishing the manuscript. All scientific content, experimental design, and results were produced by the authors.

\bibliographystyle{IEEEtran}
\bibliography{mybib}

\end{document}